# MIMO: State of the Art, and the Future in Focus

Mboli Sechang Julius

*Abstract*-Antennas of transmitters and receivers have been manipulated to increase the capacity of transmission and reception of signals. Using many elements in antennas to shape beams and direct nulls in a particular point for optimum signal transmission and reception has over decades, had tremendous positive influence in received power and signal to noise ratio (SNR). However, since the antenna elements manipulation can be done both at base station and device terminal, it gives rise to an important method of using several antennas to put and obtain signals to and from space with increased capacity. This principle is termed Multiple-input and Multiple-output (MIMO).

This paper discusses application of MIMO in the state of the art and next generation of wireless systems (5G). It also discusses four models of MIMO; SISO, SIMO, MISO and MIMO, considering three method of combing the signals from multipath propagations; Selection combining (SC), Equal gain combing (EGC) and maximum ratio combining (MRC). Spatial diversity and spatial multiplexing are also discussed as form of MIMO. Finally, Massive or Hyper MIMO which is a new method of increasing transmission capacity by very large scale for fifth generation of wireless system is discussed with its challenges and opportunities.

*Key terms*-Diversity combining techniques, spatial multiplexing, channel state information (CSI). Massive MIMO

1. INTRODUCTION

In radio communications, antennas play a critical role and their capabilities have been significantly improved within the past decades [1]. A typical antenna for today's wireless communication is technically a lone radome that usually contained multiple elements which can be manipulated to steer nulls and beams of the antenna to desired directions[1]. When an antenna is used to send or input signals into space, it is technically referred to as input device for the space and if it is used to receive signal from space, it is referred to as output device. It is this impression that gives rise to the term Multiple Input, Multiple Output (MIMO) which is when multiple antennas are used at both systems involved in wireless communications.

Bell Labs patented MIMO in 1984 [2] and since then, its conFigureuration has been extensively worked upon to improve the reliability of transmission and reception. MIMO is used on both downlink and uplink channels. For instance, when base station (BS) and user equipment (UE) are both equipped with single antenna each, it is conventionally referred to as single input, single output (SISO). In this system, the well-known Shannon formula gives the capacity as;

$$C = B\log2 (1 + SNR) \qquad (1)$$

In between SISO and MIMO lies smart antennas (Antenna diversity), which is either transmit diversity, multiple-Input, Single-Output (MISO), or receive diversity, Single-Input, multiple-Output (SIMO) [1, 3, 4].

MIMO systems can either be single-user or multiuser channels. If a transmitter uses *n* antennas while a receiver uses *m* antennas, there are then $[m \times n]$ ways of conFigureuring the system [5]. When the emitters and sensors are respectively collocated on one transmitter and one receiver which is coordinated at both ends, the system is termed "Single User MIMO (SU MIMO)" or "point to point" [6]. In this system, the receiver tries to pick up the "n" input signal(s) simultaneously. A multi user MIMO (MU MIMO) consists of multiple access and broadcast channel.

A multiple access channel is a multipoint-to-point channel where multiple users or transmitters from different geographical locations access the same source with very little or no coordination. A good example of this system is the uplink channel in cellular networks. On the other hand, MIMO broadcast channel is typically a point-to-multipoint system where users or receivers which are geographically separated receive distinct messages that are sent simultaneously by a lone transmitter which may have one or more coordinated emitters.

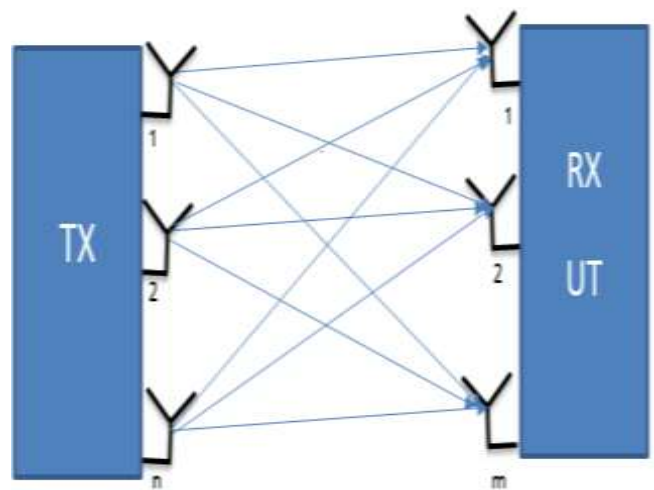

**Figureure 1: SU MIMO**

Figure 1. SU MIMO

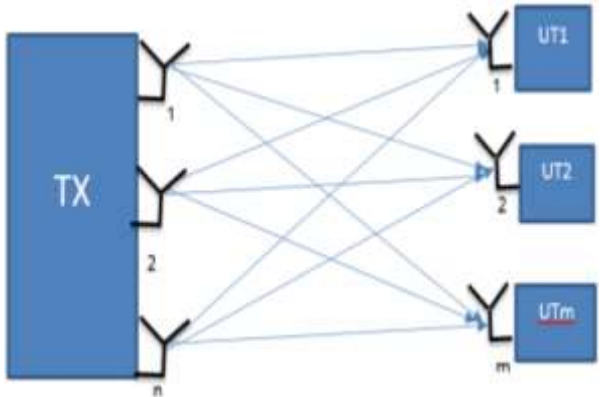

**Figureure 2: MU MIMO**

The downlink channel in cellular network is a good example. The Shannon formula for calculating the channel capacity in MIMO system is

$$C = \log_2 \det(1 + \frac{1}{\sigma_n^2} H R_x H^H) \qquad (2)$$

## 2. Spatial Diversity and Spatial Multiplexing

Spatial diversity basically makes use of multiple antennas in wireless transmission where the antennas has similar characteristics and physically separated [1]. This is usually done when the incoming signal incidence is considered, so that the effect of co-channel interference is mitigated. The use of cells and sectors in cellular networks like UMTS takes advantage of spatial diversity which helps in efficient use of the spectrum among multiple users. MISO and SIMO are essential transmit and receive diversity respectively. Diversity only makes the transmission more robust but does not increase the data rate of the transmitted signals [1, 4].

Spatial multiplexing is a method that transmits data streams or combined data for separate antennas to take advantage of the available space dimension so that it is reused. Spatial multiplexing MIMO systems have great advantage over conventional MIMO systems in that instead of single data transmission, data streams are transmitted to make the transmission more robust as opposed to ordinary spatial diversity, it is intended to increase the data rate [1]. The concepts of spatial transmit diversity and spatial multiplexing raises an interesting question; how are these signals combined and demodulated?

## 3. Combining Techniques

The existence of transmit diversity needs receive diversity as well, it is even required that the number of receive antennas equals or exceeds the transmit antennas if the MIMO system must be effective. There are mainly three techniques of combing the signals; selection combining (SC), equal gain combining (EGC) and maximum ratio combing (MRC) [1, 3, 7].

### 3.1 Selection Combining (SC)

In SC, all the signals received are first sampled and the largest is sent to demodulator for further process. This method is very easy to implement but unfortunately is not effective as it does not make simultaneous use of all the signals. This is done by simply comparing the various SNRs of the signals and eventually selecting the signal with highest SNR. SC is the cheapest method and does not require any additional RF receiver chain though in real life situation, the system only pick the strongest signal as it may be difficult to measure the SNRs individually [1, 7].

### 3.2 Maximum Ratio combining (MRC)

In MRC, the various individual signals from the different branches are co-phased and weighted with respect to their individual SNRs and then summed to obtain the output. MRC gives an average SNR output which is the same as summing the average SNRs of the several branches if it is assumed that all the branches have equal average SNR. The output SNR will be acceptable at receiver since it probably gives the largest SNR. A typical result or sum from MRC will be like the black curve in Figure.6

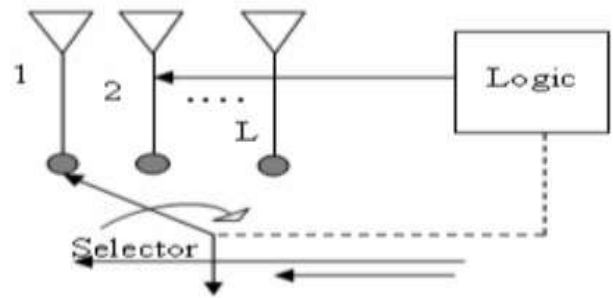

**Figure 3: Selection Combining (SC)**

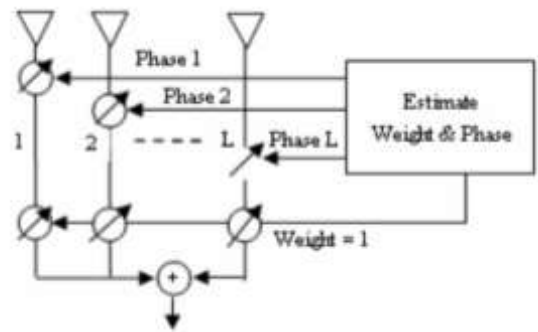

**Figure 4: Maximum Ratio combining (MRC)**

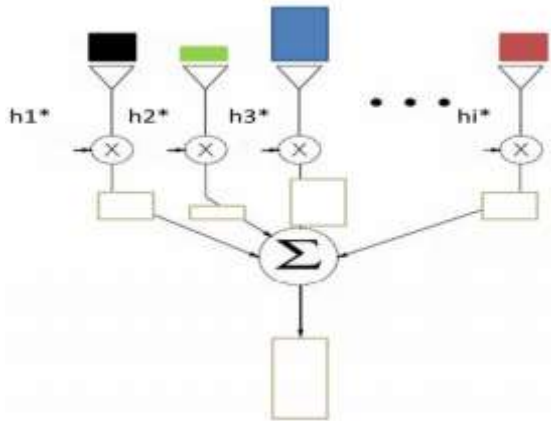

**Figure 5:** Maximum Ratio combining (MRC), Signals Addition

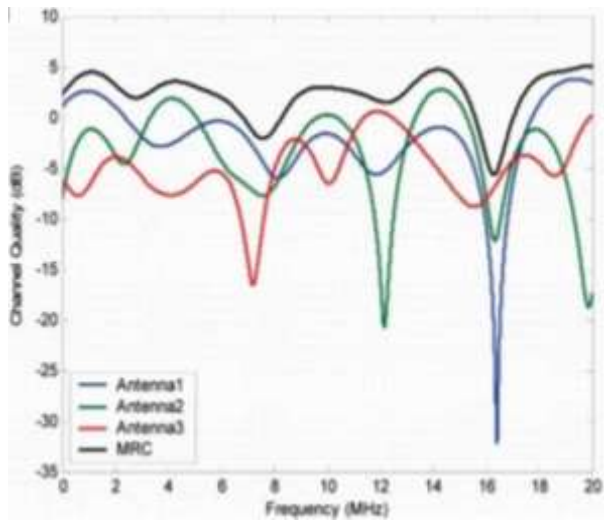

**Figure 6:** Maximum Ratio combining (MRC) output sample.

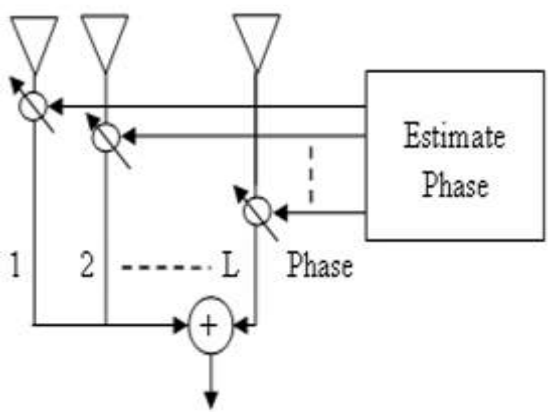

**Figure 7:** Equal Gain combining (EGC)

*3.3 Equal Gain combining (EGC)*

EGC works in similar manner as MRC but the difference is that the weights of the individual SNRs from the branches are all made unity. It is still possible to obtain an SNR that is acceptable from the various SNRs which may have been unacceptable. In comparison to MRC, EGC is less complex and its performance is marginally less effective than MRC.

### 4. STATE OF THE ART

It is now established that MIMO increases data rate, makes transmission more robust, lowers the bit error rate, increases coverage area and improves position estimation. The downside of MIMO is mainly its complexity and the limitation of meeting the ever increasing and challenging demand of data transmission [2].

In the past, second generation technology like GSM used shared transmission channel that users were allocated by dividing time and the available frequency, that is frequency division multiple access and time division multiple access (FDMA/TDMA) and third generation system like the UMTS used code with time and frequency, that is, code division multiple access (CDMA) in two modes; either frequency division duplex (FDD) or time division duplex (TDD) [8]. But now in LTE-A (long term evolution advance), MIMO is used in both uplink and downlink channels in different conFigureurations [2, 9]. When 4 antennas on the eNodeB transmit to 2 antennas in the UE, it is a $4\times2$ MIMO downlink channel, and if 4 antennas on the UE transmit to 4 antennas on the eNodeB, then it is a $4\times4$ MIMO uplink channel.

For downlink, release 8 and 9 of LTE allows up to $4\times4$ MIMO while $2\times2$ MIMO is supported in the uplink [8]. This conFigureuration is modified to $8\times8$ MIMO for downlink channel while uplink channel is increased to $4\times4$ MIMO. Downlink with 8 antennas for both eNodeB and UE (needs more power and large size) is very sophisticated and costly design. This implies that this technology may be dropped in the near future [8].

If an $8\times8$ MIMO system implements 20 MHz carrier as described above, its throughput will be about 600 Mbps [8]. With this throughput, if it is assumed that an $8\times8$ MIMO system deploys the aggregation of $5\times20$ MHz carriers in LTE-A, then maximum throughput will be 600 Mbps $\times$ 5 carriers = 3 Gbps, but it's only a theoretical value though. MIMO system is used in two methods in 3GPP (Third Generation Partnership Project); SU-MIMO and MU-MIMO. Any cell can be in any of these modes dynamically switching but not in both methods simultaneously [2, 8].

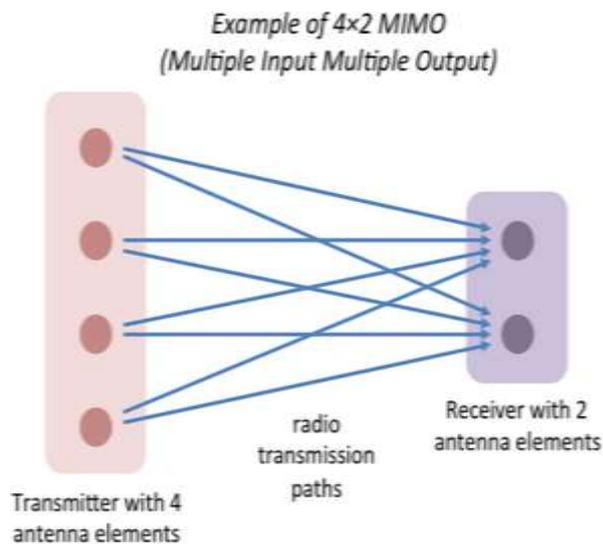

Figure 8: Illustration of 4×2 downlink MIMO. [10]

An $8\times 8$ SU-MIMO system is capable of delivering peak data rate for just one user but as mentioned earlier this technology will likely not be used because it is required that the several layers are not correlated spatially and that both the eNodeB and UE make use of 8 transmitting and receiving antennas respectively [8]. This technology will also mean less effectiveness since most applications will not actually need high data rates or deliver high data rates. Therefore, MU-MIMO technology will actually be a better option in this case. All LTE-A enabled-UEs can be used to exploit MU-MIMO since they are likely to be in different physical locations and so the MIMO layers are spatially separated, promising higher throughput in a cell or sector than that of SU-MIMO [2, 8].

The LTE downlink channel uses spatial multiplexing in two loops; either closed loop or open loop depending on whether there is feedback from the UE or not [2, 8, 10]. It is a closed loop when the channel state information (CSI) is fed back to the eNodeB and it is an open loop where there is no feedback. The transmitter would normally use the CSI to recompense loses in the channel in such a way that the receiver gets the best quality of service as much as possible. However, this leads to two issues; the feedback loop can take up high bandwidth thereby affecting its efficiency and CSI delivery delay is also a major problem when CSI changes rapidly.

The challenge of CSI delivery delay is solved by using codebooks [8] in LTE-A, but the problem of bandwidth efficiency remains a major challenge since radio resource is limited. For instance, in $2\times 2$ MIMO system, up to 4 reports channels are expected to be delivered while 16 reports are required for a $4\times 4$ MIMO system and formations of 64 channels are needed for $8\times 8$ MIMO system that is supported in LTE-A. The use of codebook which is simply a set of predefined matrices with each standing for a kind of transmission parameters for a particular channel, is limited in size. CSI problem is much simpler to solve in TDD mode since it employs same channel for both uplink and downlink channel [2, 8, 10].

Generally, it is a good idea to employ spatial multiplexing, when good channel conditions exist, but when the channel conditions are poor, then an alternative method should be transmit diversity. This is true because for successful performance, spatial multiplexing will definitely need high quality radio channel but if the channel is weak, then multiple antennas can be employed to proffer solution for high throughput [8].

The uplink channel in LTE-A uses MIMO that is a bit different from the downlink MIMO. For uplink channel, first thing to consider is size of UE which usually should be very small and the desire to make it even smaller and flexible is ever on the increase. Release 11 of LTE support maximum number of 4 antennas due to their small size in comparison to eNodeB and since they are hand hell devices, techniques like beamforming in uplink channels is impracticable [8].

## 5. NEXT GENERATION MIMO

As everyone is already talking of fifth generation technology (5G), there is also need for evolution in MIMO since it still suffers some challenges at this present generation. The new research area already in progress for this 5G technology is termed as massive MIMO. Simply put as the technology that will furnish base stations with very high number of antennas (hundreds or even thousands) in comparison to conventional MIMO [8, 11, 12]. Massive MIMO may also be called very large MIMO, ARGOS, full dimension MIMO or hyper MIMO [11, 13].

5G networks are expected to support a very dense system with large numbers of devices that will support machine-to-machine and machine-to-person communications [5, 14]. Therefore, the frequencies range for massive MIMO is expected to go from centimetre/millimetre range, from 6 GHz to about 100 GHz to support these ever craving data consumers. The traffic will likely increase tremendously. One billion new mobile subscribers expected to join the existing 3.6 billion to make it 4.6 billions by 2020 [15], and to cope with this increase, Massive MIMO might need to proffer up to 100 Mbit/s data speed and peak data speed of about 10 Gbit/s [14].

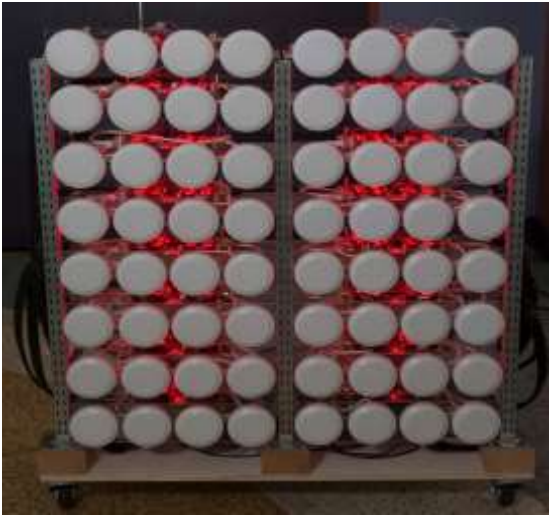

Figure 9: Sample Base station with many antennas [16].

Major advantages include:

i. Increase in capacity by a factor of 10 or more.
ii. Very high increase in data rate due to large number of antennas leading to high transmits diversity.
iii. Low and cheap power components can also be used to build Massive MIMO so that its output power lies in the mili-watt range.
iv. Efficiency will largely be increased due to the fact that beams and nulls from base station will be steered to where they are needed, thereby avoiding wastage and interference [17].
v. Latency can be greatly reduced in the open air interface because fading dips can be avoided with beamforming.

The main aim of the coding techniques in massive MIMO is to be able to purely separate the distinct users' data streams at the receiver by employing optimal maximum likelihood detector, decision feedback detectors and linear detectors[16].

However, with so much anticipated benefits of massive MIMO, there are limitations and research challenges. The limitations include pilot contamination, channel reciprocity, non-channel state information (CSI) at the transmitter operation and physical separation between base stations and users. If UEs with line of sight (LOS) are close to the base station, interference will occur due to high correlation. The number of antennas will also be limited at certain level where additional antennas will not make any change again [5, 11].

Since the performance of wireless communication systems largely depends and in fact, is governed by the wireless channel environment, new and hot research areas is springing up in in MIMO or topics connected to it. All new technologies being developed daily from the several applications of Internet of Things (IoT) and Big Data depend either directly or indirectly on high and reliable data rate which can be possible via Massive MIMO as already discussed. The analysis of wireless communication systems is always difficult due to the fact that the wireless channels employed is always unpredictable and dynamic. This and many challenges are opening up research opportunities in wireless communications.

6. CONCLUSION

This paper gave a general overview of MIMO as a system that started with just SISO and went all the way to MIMO. It related how transmission all started with just a single line and progresses onto multiple channels. The desire for more efficiency and quality of service in the communication industries led us to MIMO and now Massive MIMO is on the way to even make the system more suitable for 5G networks. This is possible because generations of networks has been evolving every decade, so it is normal to expect 5G in 2020.